\documentclass[aps,prb,twocolumn,showpacs,amsmath,amssymb]{revtex4}

\usepackage{graphicx} 
\usepackage{dcolumn} 
\usepackage{bm} 
\usepackage{epsfig}
\usepackage{amsmath}
\usepackage{amssymb}
\usepackage{hyperref}
\usepackage{color}
\usepackage{ulem}
\begin{document}
\newcommand{\bsy}[1]{\mbox{${\boldsymbol #1}$}} 
\newcommand{\bvecsy}[1]{\mbox{$\vec{\boldsymbol #1}$}} 
\newcommand{\bvec}[1]{\mbox{$\vec{\mathbf #1}$}} 
\newcommand{\btensorsy}[1]{\mbox{$\tensor{\boldsymbol #1}$}} 
\newcommand{\btensor}[1]{\mbox{$\tensor{\mathbf #1}$}} 
\newcommand{\tensorId}{\mbox{$\tensor{\mathbb{\mathbf I}}$}} 
\newcommand{\be}{\begin{equation}}
\newcommand{\ee}{\end{equation}}
\newcommand{\bea}{\begin{eqnarray}}
\newcommand{\eea}{\end{eqnarray}}

\title{Slow light in semiconductor quantum dots: effects of non-Markovianity and correlation of dephasing reservoirs}

\author{D. Mogilevtsev$^{1,2}$, E. Reyes-G\'omez$^{3,4}$, S. B. Cavalcanti$^{5}$, and L. E. Oliveira$^{4}$ }

\affiliation{$^1$Centro de Ci\^encias Naturais e Humanas,
Universidade Federal do ABC, Santo Andr\'e,  SP, 09210-170 Brazil \\
$^2$Institute of Physics, NASB, F. Skarina Ave. 68, Minsk, 220072, Belarus \\
$^3$Instituto de F\'{i}sica, Universidad de Antioquia UdeA, Calle 70 No. 52-21,
Medell\'{\i}n, Colombia \\
$^4$Instituto de F\'{i}sica, Universidade Estadual de Campinas - Unicamp, Campinas - SP, 13083-859, Brazil \\
$^5$Instituto de F\'{\i}sica, Universidade Federal de Alagoas, Macei\'{o} - AL, 57072-970, Brazil}

\begin{abstract}
A theoretical investigation on slow light propagation based on
eletromagnetically induced transparency in a three-level
quantum-dot system is performed including non-Markovian effects
and correlated dephasing reservoirs.  It is demonstraonated that
the non-Markovian nature of the process is quite essential even
for conventional dephasing typical of quantum dots leading to
significant enhancement or inhibition of the group velocity
slow-down factor as well as to the shifting and distortion of the
transmission window. Furthermore, the correlation between
dephasing reservoirs may also either enhance or inhibit
non-Markovian effects.
\end{abstract}


\date{\today}

\maketitle

\section{Introduction}

Slow light group velocity propagation has revealed itself to be a
key stone in the construction and design of variable delay lines
which are of great importance in the synchronization of optical
signals and signal buffering in all-optical communication systems.
One of the most promising approaches is the one implementing
electromagnetically induced transparency (EIT). \cite{eit_review}
It has been experimentally demonstrated that in EIT schemes it is
possible to obtain a slow-down factor of $10^7$ in gases, such as
Rb vapor \cite{gas} and cold cloud of sodium atoms, \cite{sodium}
or solid-state systems as Pr doped \cite{solid} Y$_2$SiO$_5$.
However, the transmission bandwidths obtained in these systems are
too narrow \cite{buffers} (about 50-150 KHz) for optical buffers.
Semiconductor structures such as quantum wells
\cite{wells,ultrawells} and quantum dots
\cite{cu,kim,peng,mark,bare} may offer much broader transmission
bandwidths (about a few GHz) at the cost of a smaller slow-down
factor and with EIT buffers operating at room temperature.
\cite{room,su} To open up possibilities in light control, one may
combine EIT semiconductor structures with other systems capable of
slowing light, i.e., photonic crystals \cite{lavr,krauss,baba} or
coupled quantum-dot heterostructures. \cite{braz}

One feature of the semiconductor nanostructures such as quantum
dots and wells is the interaction with the substrate host. For
example, quantum dots interact with phonons or with carriers
captured in traps in the vicinity of the quantum dot. The
influence of the surroundings leads to dephasing and to energy
loss of the semiconductor nanostructure. The most common
dephasings and energy losses may be described by Markovian master
equations usually written down in the so-called Lindblad form
\cite{Lindblad} (see, for example, the recent works by Colas
\textit{et al} \cite{Colas} and Marques \textit{et al}
\cite{Marques}). Markovianity arises when correlations of
dephasing or dissipative reservoirs rapidly decay on the typical
time-scale of the nanostructure dynamics. However, it is well
known that the dephasing process in solids is quite often of a
non-Markovian nature as suggested in experimental \cite{dewoe} and
theoretical \cite{kilin} studies. Reservoir correlations do not
decay quickly enough and the density of reservoir states changes
significantly on the scale of reservoir-system interaction
constants and Rabi frequencies of driving fields. It is important
to note that non-Markovian dephasing in quantum dots is
responsible for phenomena such as, for instance, the damping of
Rabi oscillations and excitation-induced dephasing,
\cite{forstner,our prl,rams,monni,ulrich,nazir,hkim}
phonon-induced spectral asymmetry, \cite{hkim,hug1,lodahl,hug2}
and interference between phononic and photonic reservoirs.
\cite{calic,vuk,kau}

To the best of our knowledge, only Markovian and uncorrelated
dephasing  have been considered in semiconductor EIT systems. In
the present study, we show, for semiconductor quantum-dot systems,
that the non-Markovian nature of dephasing leads to a significant
modification of the group velocity slow-down factor in comparison
with the Markovian one. Moreover, it is demonstrated that the
absorption spectrum becomes asymmetric and the transmission window
is shifted and modified. It is interesting that, in contrast with
the damping of Rabi oscillations, \cite{forstner,our
prl,rams,monni} the changes undergone by the slow-down factor and
the transmission window are actually first-order effects on the
frequency of the driving field associated with the Rabi
oscillations. They may occur even in situations where the
driving-induced dephasing and the damping of driven Rabi
oscillations do not take place. Furthermore,  correlations between
dephasing reservoirs is also considered. Correlation between
losses were already shown to lead to a number of non-trivial
effects in the dynamics of open systems.
\cite{ekert,lidar,valera,our chain} In this respect, well-known
decoherence-free subspaces result from different couplings of the
system with the same reservoir, and may be used for avoiding
decoherence of the quantum states. \cite{ekert,lidar} Correlated
losses may be exploited to create nonlinear loss in deterministic
non-classical states generation \cite{valera} and to produce
excitation flow like heat while retaining coherence. \cite{our
chain}

The outline of the present study is as follows. In Section II we
introduce the model of the three-level quantum dot in a ladder
configuration under the action of both a strong pump and of a weak
signal optical fields. The master equation and corresponding
stationary solutions are presented in section III, whereas the
susceptibility, absorption and slow-down factor are given in
Section IV.  In Section V, the influence of non-Markovian effects
on the slow-down factor and absorption as well as some examples
are analyzed. Also, a comparison with the Markovian case is
demonstrated. Finally, discussion and conclusions are presented in
Sections VI and VII, respectively.

\begin{figure}[t]

\includegraphics[width=0.95\linewidth]{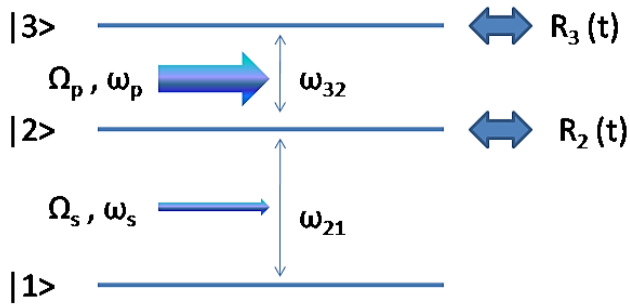}

\caption{Pictorial view of a three-level quantum dot driven by a strong pump field with frequency $\omega_p$ and Rabi frequency $\Omega_p$. $R_{2}$ and $R_{3}$ denote dephasing reservoirs coupled to the corresponding levels described by states $\vert 2\rangle$ and $\vert 3\rangle$, respectively. The signal field with frequency $\omega_s$ and Rabi frequency $\Omega_s$. Dipole allowed transition frequencies are denoted by $\omega_{32}$ and $\omega_{21}$.}
\label{fig1}
\end{figure}

\section{The three-level quantum-dot model}

The experimental setup needed to observe EIT involves two highly coherent optical fields interacting with a three level system in various schemes. Here, we are interested in a quantum dot model and therefore we choose a ladder scheme for EIT as depicted in Fig. \ref{fig1}: a weak signal field is tuned near resonance with the $\vert 1     \rangle \rightarrow \vert 2 \rangle$ transition, while a strong pump quasi-classical field is tuned with the  $\vert 2 \rangle \rightarrow \vert 3 \rangle$ transition. We denote by $\omega_s$ and $\omega_p$ the optical frequencies of the signal and pump fields, respectively, and  $\Omega_s$, $\Omega_p$ the Rabi frequencies associated with the signal and pump fields. Such a scheme was considered by Kim \textit{et al.} \cite{kim} for a strained GaAs-InGaAs-InAs quantum-dot system. We assume that $\omega_{21}$ and $\omega_{32}$ are sufficiently different so that the driving by the $\omega_p$ strong pump field (with $\Omega_p$ Rabi frequency) only occurs between upper states $\vert 2\rangle$ and $\vert 3\rangle$. Similarly, the $\omega_s$ weak signal field (with $\Omega_s$ Rabi frequency, $\vert \Omega_s\vert \ll\vert \Omega_p\vert $) acts only between states $\vert 1\rangle$ and $\vert 2\rangle$. Also, to illustrate the effects of the non-Markovian character of the dephasing process,  we assume that energy loss occurs on times far exceeding the typical dephasing time of the quantum dot (e.g., in the study by Kim \textit{et al.}, \cite{kim} for a GaAs-InGaAs-InAs quantum-dot system, dephasing times were about several tens of picoseconds at temperatures of  50 K - 80 K, whereas energy loss occurred on the scale of several nanoseconds). In the rotating-wave approximation and interaction representation, the three-level quantum dot scheme depicted in Fig. \ref{fig1} may be described by the following generic Hamiltonian
\bea
\label{ham}
V(t)&=&\hbar[\Delta_s+\Delta_p+R_3(t)]\sigma_{33}+\hbar[\Delta_s+R_2(t)]\sigma_{22} \nonumber \\
&+&\hbar\frac{\Omega_p}{2}(\sigma_{32}+\sigma_{23})+\hbar\frac{\Omega_s}{2}(\sigma_{21}+\sigma_{12}),
\eea
where, for simplicity, we assume $\Omega_{p}$ and $\Omega_{s}$ to be real, operators $\sigma_{kl}=\vert k \rangle \langle l \vert $ ($k,l=$1, 2, 3), $ \vert k\rangle$ describes the $k$-th state of the system, and detunings are $\Delta_s=\omega_{21}-\omega_s$ and $\Delta_p=\omega_{32}-\omega_p$. The Hermitian operator $R_k = R_k(t)$ describes the dephasing reservoirs influencing the transition to the $k$-th state. We notice that one may exclude an action of a dephasing reservoir on the lower state by using $\sum\limits_{k=1}^3\sigma_{kk}=1$. Thus, each operator $R_k$ contains variables of two reservoirs (see, for example, the study by Kaer \textit{et al.} \cite{lodahl}). Here we do not specify the exact nature of the dephasing reservoirs. It might include all types of pure dephasing encountered in quantum dots beyond the independent linear boson model commonly used to describe the phonon interaction with the dot, such as effects of possible quadratic coupling of phonons to the dot \cite{mul} or phonon-phonon scattering. \cite{rudin} We require only the existence of the quantities
\bea
\label{fourier cor }
D_{kl}^{\pm}(\delta)=\lim \limits_{t \rightarrow \infty} \int \limits_0^t
d \tau K^{\pm}_{kl} (t,\tau) e^{i\delta \tau}
\eea
for any real $\delta$, where $K_{kl}^+(t,\tau)=\langle R_k(t)R_l(\tau)\rangle$ and $K_{kl}^-(t,\tau)=\langle R_k (\tau) R_l (t) \rangle$. For realistic many-system reservoirs one usually has $\langle R_k (x) R_l(y) \rangle \rightarrow 0$ for $\vert x-y \vert, \,x, \,y \rightarrow + \infty$ and the quantities \eqref{fourier cor } do exist, defining, for example, asymptotic decay rates obtained with the time-convolutioness master equation. \cite{BP}

\section{The master equation}

There is a strong pump field driving transition between states $\vert 2\rangle$ and $\vert 3\rangle$ of the  three-level quantum dot. To derive a master equation accounting for effects of the strong dot-field interaction, it is necessary to transform to a dressed picture with respect to the operator
\be
\label{Hdress}
V_d=\hbar\Delta_p\sigma_{33} + \hbar\frac{\Omega_p}{2} (\sigma_{32}+\sigma_{23}).
\ee
The dressed Hamiltonian is
\bea
\label{Hdressed}
\mathbf{V}(t)&=& \hbar [\Delta_s+R_3(t)]{\bf S}_{33}(t) + \hbar [\Delta_s + R_2(t)]\mathbf{S}_{22} (t) \nonumber \\
&+& \hbar \frac{\Omega_s}{2}[\mathbf{S}_{21}(t)+\mathbf{S}_{12}(t)],
\eea
where the dressed operators are $\mathbf{S}_{kl}(t) = U^{\dagger} (t) \, \sigma_{kl} \, U(t)$, and $U(t) = \exp (-i\, V_d \, t / \hbar)$. Thus, for the population operators, one has \cite{our prl}
\begin{subequations}
\label{popdressed}
\bea
\label{popdressed-a}
\mathbf{S}_{33}(t)&=&S_{33}(t)\sigma_{33}+S_{23}(t)\sigma_{23}+S_{32}(t)\sigma_{32} \nonumber \\
&+&S_{22}(t)\sigma_{22}
\eea
and
\be
\label{popdressed-b}
\mathbf{S}_{22}(t) =\sigma_{33}+\sigma_{22}- {\bf S}_{33}(t),
\ee
\end{subequations}
where
\begin{subequations}
\label{sdressed}
\be
\label{sdressed-a}
S_{33}(t)=\frac{1}{2}\left [ 1 + c^2 + s^2 \cos(\Omega_Rt) \right ],
\ee
\be
\label{sdressed-b}
S_{22}(t)=\frac{s^2}{2}\left [ 1-\cos(\Omega_Rt) \right ],
\ee
and
\be
\label{sdressed-c}
S_{23}(t)=\frac{s}{2} \left \{ c [1 - \cos(\Omega_Rt) ]+ i \sin(\Omega_Rt) \right \},
\ee
\end{subequations}
with $\Omega_R^2=\Delta_p^2+\Omega_p^2$, $c=\Delta_p/\Omega_R$, $s=\Omega_p/\Omega_R$ and $S_{32}(t)=S_{23}^*(t)$.

Notice that here we assume low temperatures, so it is not necessary to consider multi-phonon processes and to implement polaron transformations to account for them. For low temperatures one may use the time-convolutioness master equation to describe the dynamics of the dot dressed by the driving field \cite{lodahl} and derive in a standard manner the master equation for the ${\bar \rho}(t)$ \, dressed density matrix of the dot. \cite{BP} Up to second order on the interaction involving reservoirs, one arrives at the following equation for the dressed density matrix averaged over the reservoirs
\bea
\label{master dressed}
\frac{d}{dt}{\bar \rho}(t)&=&-i \left [ \Delta_s \{ \mathbf{S}_{22}(t)+\mathbf{S}_{33}(t) \}, {\bar \rho}(t) \right ] \nonumber \\
&+& i \left [ \Omega_s \{ \mathbf{S}_{21}(t) + \mathbf{S}_{12}(t) \},{\bar \rho}(t) \right ] \nonumber \\
&-& \int\limits_{0}^td \tau \left \langle \left [ V_R(t), \left [ V_R(\tau),{\bar \rho}(t) \right ]  \right ] \right \rangle,
\eea
where $V_R(t)=R_3(t)\mathbf{S}_{33}(t)+R_2(t) \mathbf{S}_{22}(t)$. Returning to the bare basis, one obtains the following equations for the off-diagonal (coherence) elements of the density matrix $\rho_{kl}=\langle k\vert \rho\vert l\rangle$,
\begin{subequations}
\label{off equations}
\bea
\label{off equations-a}
\frac{d}{dt}\rho_{13}(t)&=&-[\gamma_{3}(t)-i(\Delta_s+\Delta_p)]\rho_{13}(t) \nonumber \\
&+& \left [\nu_3(t)-\frac{i}{2} \Omega_p \right ] \rho_{12}(t) + \frac{i}{2}\Omega_s \rho_{23}(t),
\eea
and
\bea
\label{off equations-b}
\frac{d}{dt}\rho_{12}(t)&=&-[\gamma_{2}(t)-i\Delta_s]\rho_{12}(t) + \left [ \nu_2(t)-\frac{i}{2}\Omega_p \right ]\rho_{13}(t) \nonumber \\
&+& \frac{i}{2}\Omega_s[\rho_{22}(t)-\rho_{11}(t)].
\eea
\end{subequations}
Notice that the form of the equations for coherences [Eqs. (\ref{off equations-a}) and (\ref{off equations-b})]  is the same for both Markovian and non-Markovian
cases. However, a non-Markovian process  leads to time-dependent
coefficients in  Eqs. (\ref{off equations}) and also to their
dependence on the driving field. The time-dependent decay rates in
Eqs. \eqref{off equations} are
$\gamma_k(t)=\gamma_k^{M}(t)+\gamma_k^{C}(t)$, where the
$\gamma_k^{M}(t)$ represent decay rates for zero pump, \be
\label{gammar} \gamma_k^{M}(t)=\int\limits_0^td\tau
K_{kk}^-(t,\tau), \quad k=2,\,3, \ee and the driving-dependent
parts of the total decay rates are
\begin{subequations}
\label{gammat}
\be
\label{gammat-a}
\gamma_2^{C}(t) = \int\limits_0^td \tau \, S_{22}(\tau-t) \left [ K_{32}^-(t,\tau)-K_{22}^-(t,\tau) \right ]
\ee
and
\be
\label{gammat-b}
\gamma_3^{C}(t) = \int\limits_0^td\tau \, \left [ 1 - S_{33}(\tau-t) \right ] \left [ K_{23}^-(t,\tau) - K_{33}^-(t,\tau) \right ].
\ee
\end{subequations}
The time-dependent cross-coherence coupling parameters in Eqs. (\ref{off equations}) are
\begin{subequations}
\label{coupl}
\be
\label{coupl-a}
\nu_2(t)=\int\limits_0^td \tau \, S_{32}(\tau-t)\left [ K_{32}^-(t,\tau)-K_{22}^-(t,\tau) \right ]
\ee
and
\be
\label{coupl-b}
\nu_3(t)=\int\limits_0^td \tau \, S_{23}(\tau-t) \left [ K_{33}^-(t,\tau)-K_{23}^-(t,\tau) \right ].
\ee
\end{subequations}
Notice that the decay rates introduced by Eqs. \eqref{gammar} and \eqref{gammat} might include imaginary components corresponding to driving-dependent frequency shifts. Eqs. \eqref{gammar}, \eqref{gammat}, and \eqref{coupl} give a clear idea about the influence of the reservoir correlation on the dynamics. Complete correlation between reservoirs, i.e., $K_{kl}^-(t,\tau)=K_{kl}^+(t,\tau)$, $k=2,\, 3$, washes out the influence of an arbitrarily strong pump driving on the coherences dynamics. On the other hand, complete anti-correlation enhances the driving influence.

\begin{figure}[t]
\begin{tabular}{c}
  \textbf{(a)} \\
 \includegraphics[scale=0.4]{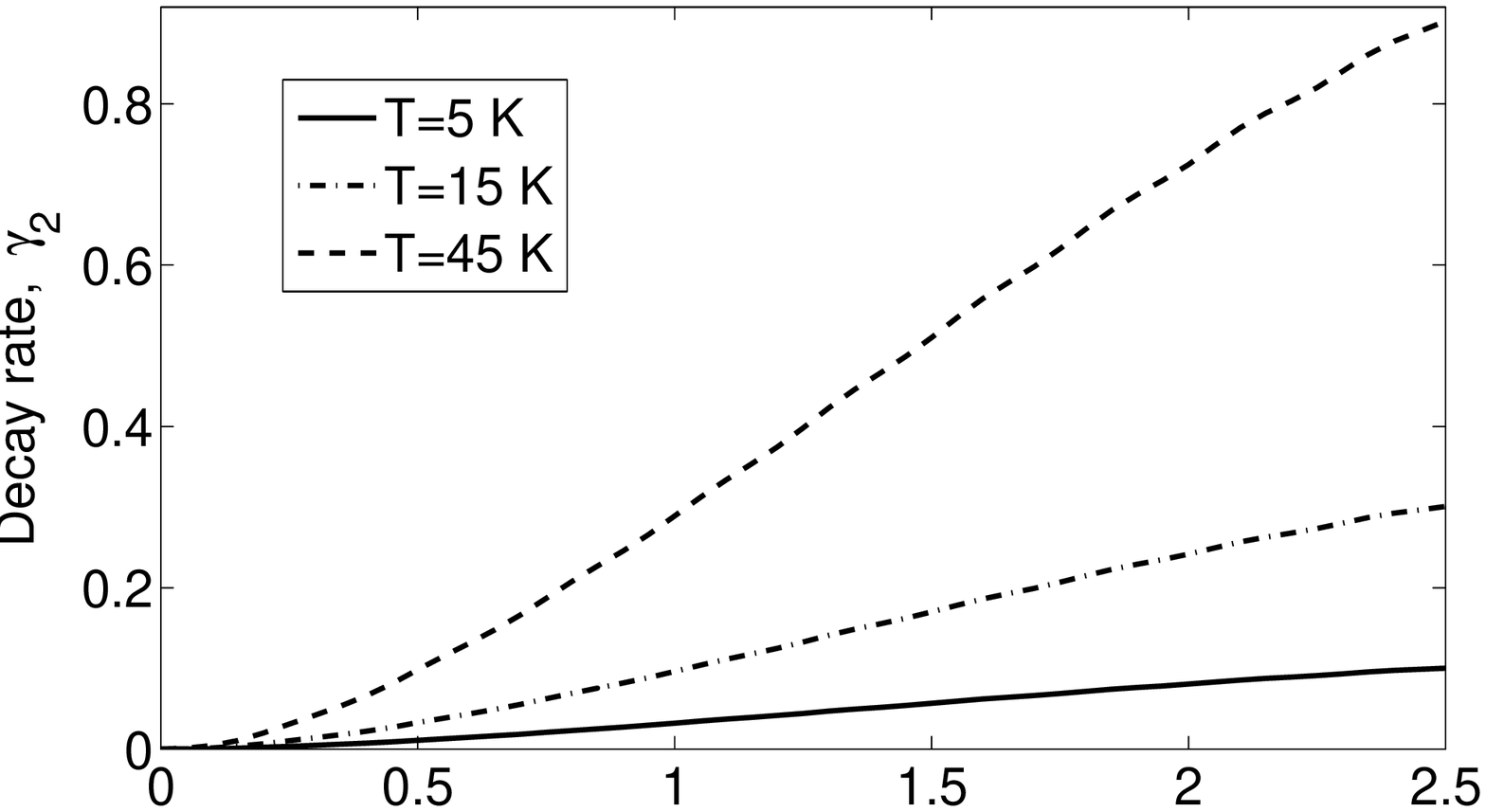}  \\
 \textbf{(b)} \\
  \includegraphics[scale=0.4]{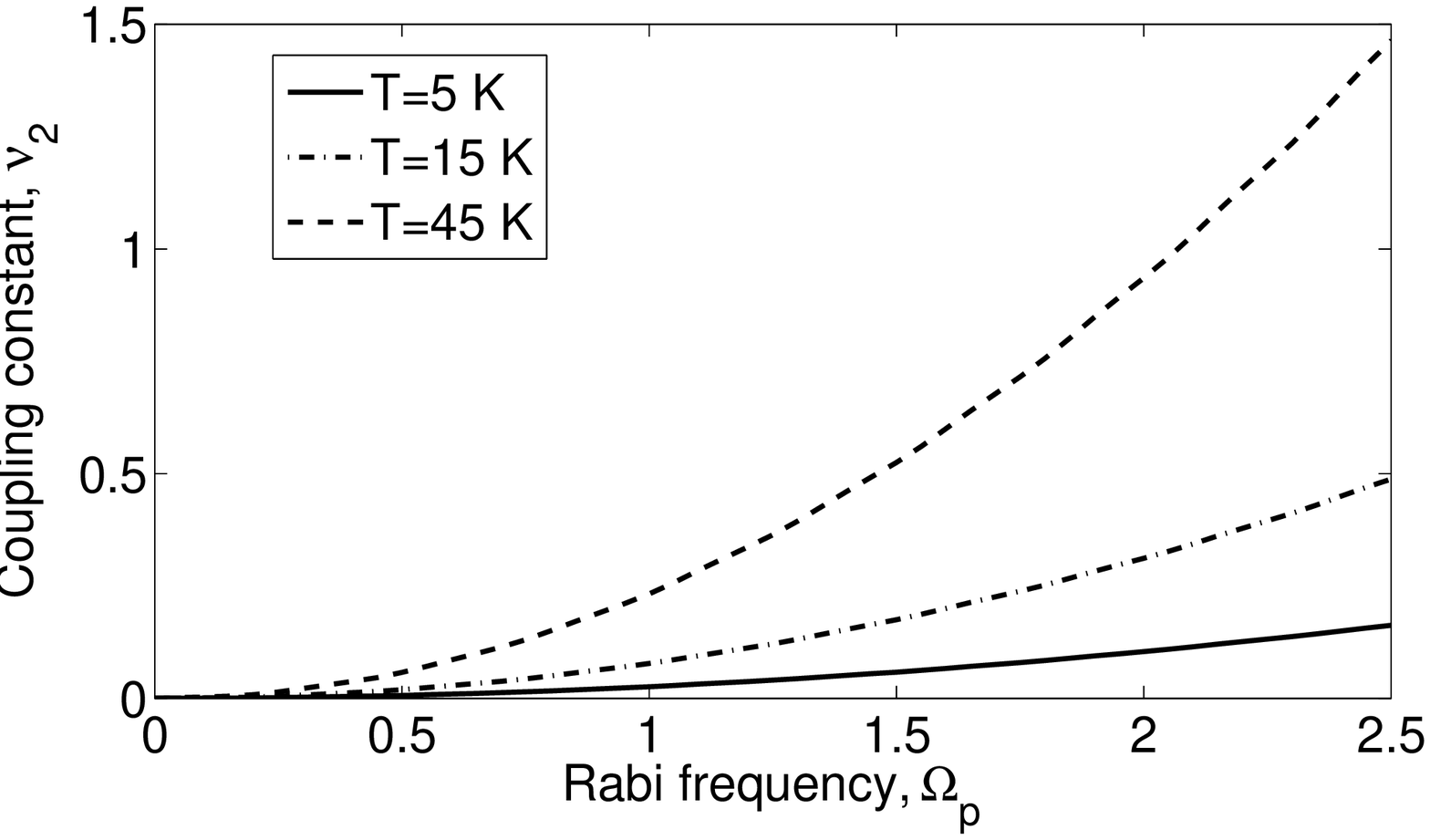}
\end{tabular}
\caption{(a) $\gamma_2$ decay rate and (b) $\nu_2$ coupling constant as functions of the Rabi frequency $\Omega_p$, according to the reservoir correlation function given by Eq. (\ref{k33}). All quantities are defined in units of the typical Rabi frequency $\Omega_0=6.6$ $\mu$eV. \cite{kim} Solid, dash-dotted and dotted lines correspond to temperatures $T=$5 K, 15 K, and 45 K, respectively.}
\label{fig2}
\end{figure}

\begin{figure}[t]
\begin{tabular}{c}
  \textbf{(a)} \\
 \includegraphics[scale=0.4]{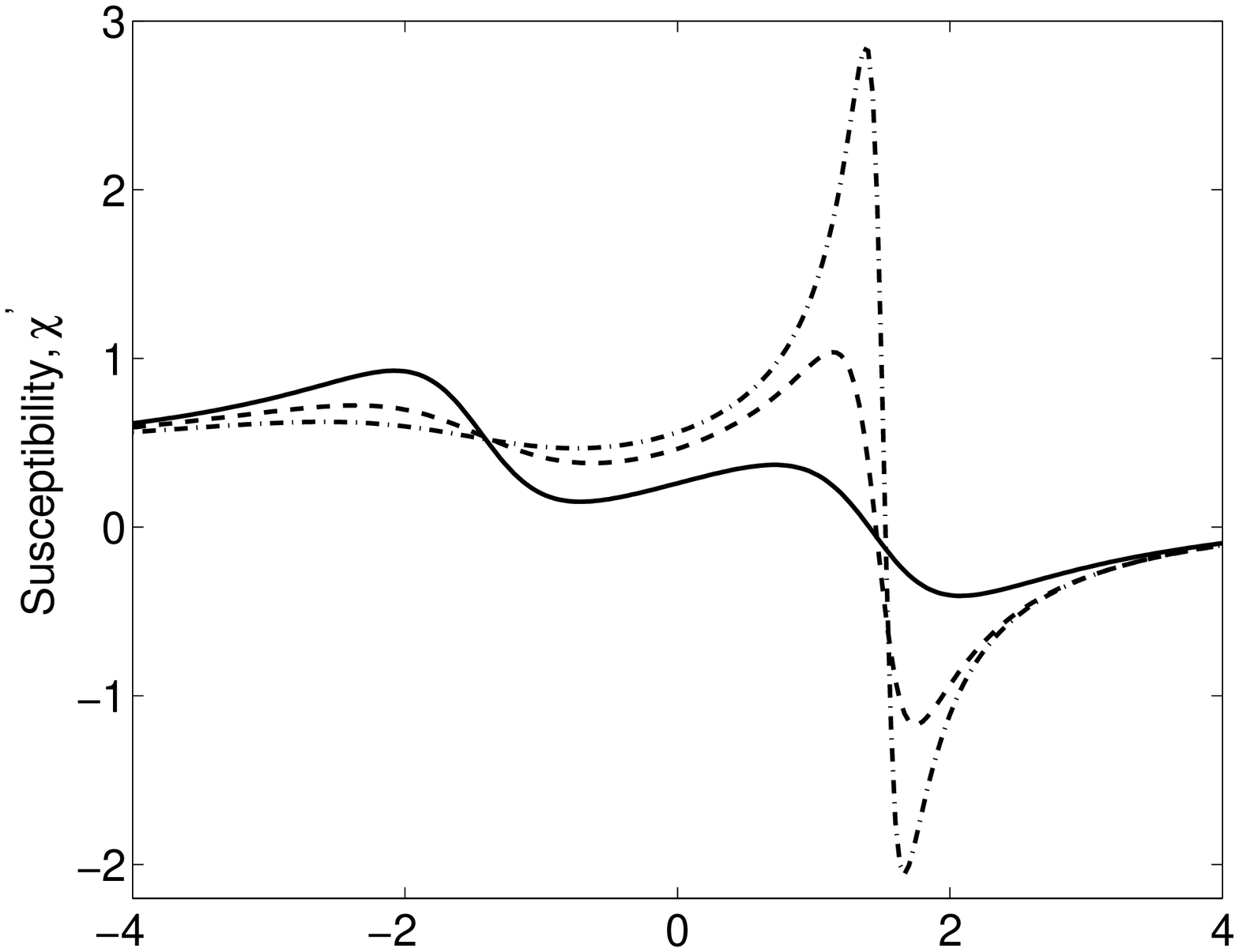}  \\
 \textbf{(b)} \\
  \includegraphics[scale=0.4]{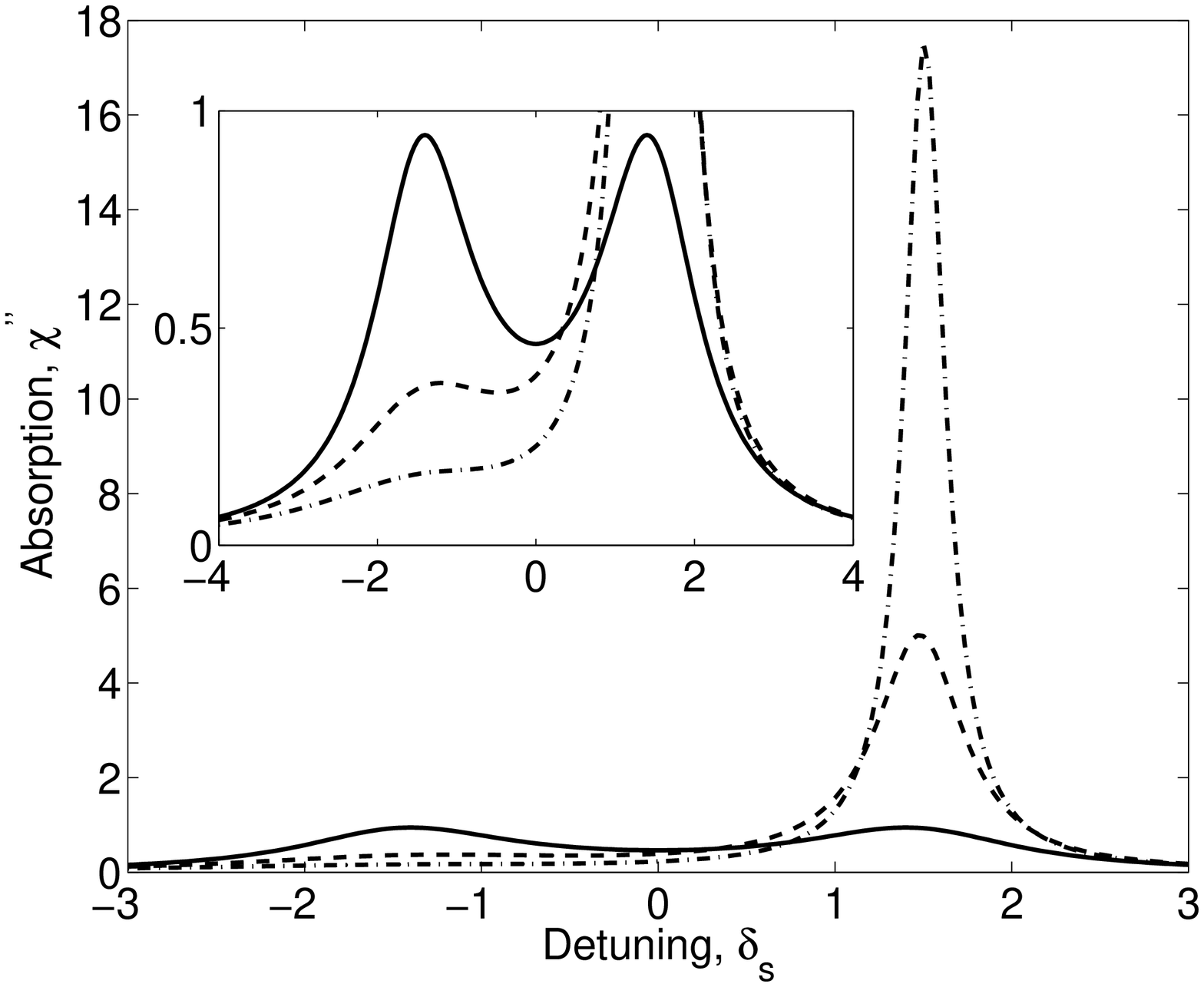} \\
\end{tabular}
\caption{(a) Normalized linear susceptibility and (b) absorption profile in the weak non-Markovian case assuming $\gamma_2$ and $\gamma_3$ independent of $\Omega_p$ and $\nu_2\approx f_2 \, \Omega_p$ [cf. Eq. (\ref{weak}) with $g_2 = g_3 = 0$]. Here we take $\gamma_3 = 2 \, \gamma_2 = 2 \, \Omega_0$, $\nu_3 = 2 \, \nu_2$, $\delta_p=0$, and $\Omega_p=3 \, \Omega_0$. All parameters are given in units of $\Omega_0=6.6$ $\mu$eV.  Solid, dashed and dash-dotted lines correspond to $f_2=0$ Markovian case, $f_2=0.1$ $\Omega_0$, and $f_2=0.2$ $\Omega_0$, respectively. Other parameters are the same as in the study by Kim \textit{et al.} \cite{kim}: $\varepsilon_{bac}=13$, $\vert \mu_{21}\vert /e=2.1$ nm, $\Gamma=0.006$, $V=8.9064 \times 10^{2}$ nm$^3$, and the wavelength of the signal field is 1.36 $\mu$m. Inset in panel (b) shows the absorption profiles in more detail.} \label{fig3}
\end{figure}

\begin{figure}[t]
\begin{tabular}{c}
  \includegraphics[scale=0.4]{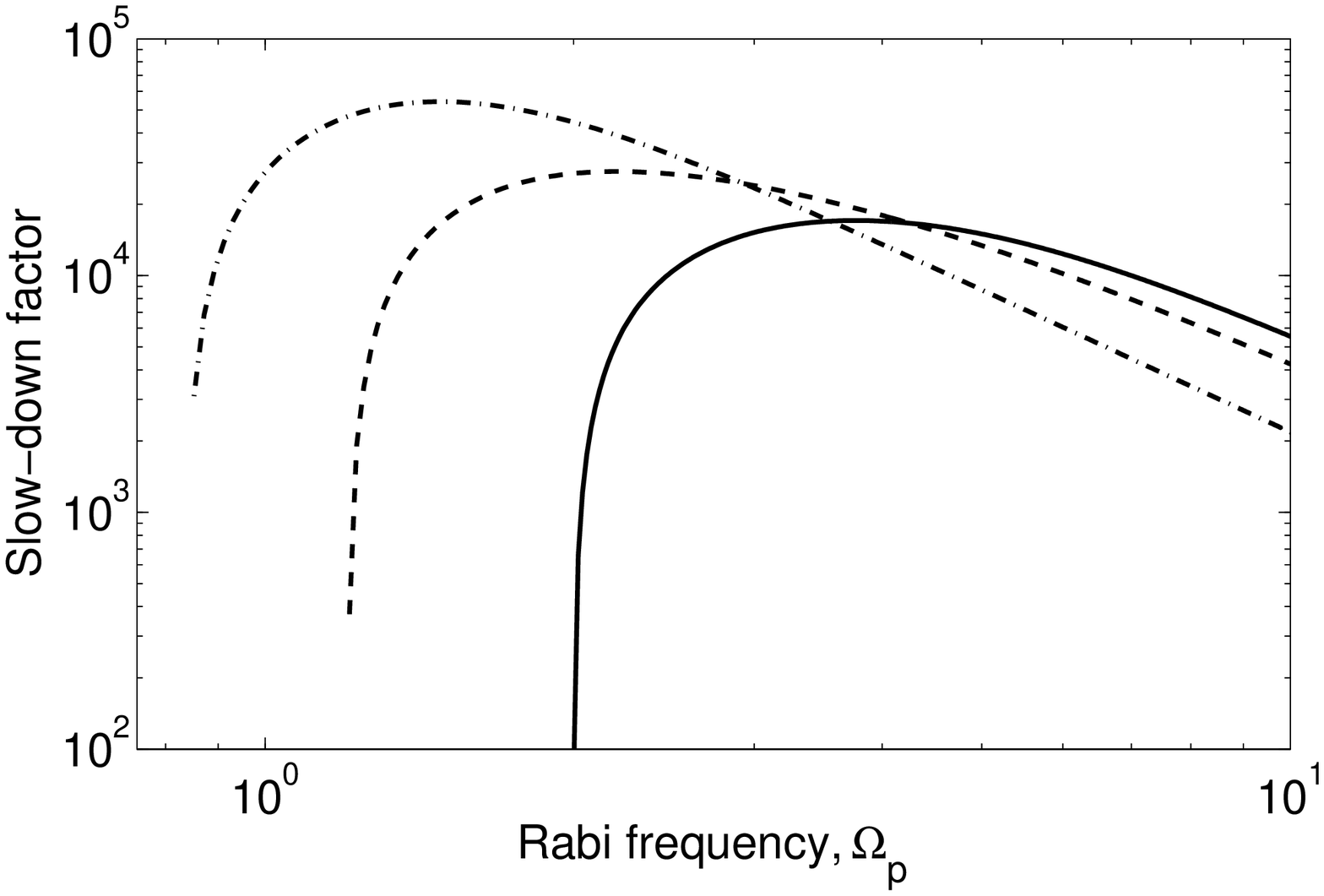}
  \end{tabular}
\caption{Slow-down factor [cf. Eq.(\protect\ref{slow})] in the weak non-Markovian case assuming $\nu_2\approx f_2 \, \Omega_p$ [$g_2 = g_3 = 0$ in Eq. (\ref{weak})] and $\delta_p = \delta_s=0$.  Solid, dashed and dash-dotted lines correspond to the $f_2=0$ Markovian case, $f_2=0.1$ $\Omega_0$, and $f_2=0.2$ $\Omega_0$, respectively. The other parameters are chosen as in Fig. \ref{fig3}.} \label{fig4}
\end{figure}

\section{Susceptibility, absorption and slow-down factor}

Now let us consider the solution of the system of Eqs. \eqref{off equations} in the $t\rightarrow\infty$ limit.
We assume that $\lim\limits_{t\rightarrow\infty}\rho_{kl}(t)\equiv \rho_{kl}^{st}$, $\lim\limits_{t\rightarrow\infty}\nu_{k}(t)\equiv \nu_{k}$, and introduce real asymptotic values of the dephasing rates
\be
\gamma_k = \mathrm{Re} \left [ \lim\limits_{t\rightarrow\infty} \gamma_{k}(t) \right ],
\ee
and modified detunings
\begin{subequations}
\be
\delta_s = \Delta_s + \mathrm{Im} \left [ \lim\limits_{t\rightarrow\infty}\gamma_{2}(t) \right ]
\ee
and
\be
\delta_p = \Delta_p + \mathrm{Im} \left [ \lim\limits_{t\rightarrow\infty}\gamma_{3}(t) \right ].
\ee
\end{subequations}
The stationary solutions for the coherence are then given by
\begin{subequations}
\label{stac1}
\be
\label{stac1-a}
\rho_{13}^{st}=\frac{(2\nu_3-i\Omega_p)\rho_{12}^{st}-i\Omega_s\rho_{23}^{st}}{2 [ \gamma_{3}-i(\delta_s+\delta_p) ]}
\ee
and
\be
\label{stac1-b}
\rho_{12}^{st}=\frac{(2\nu_2-i\Omega_p)\rho_{13}^{st}-i\Omega_s(\rho_{22}^{st}-\rho_{11}^{st})}{2(\gamma_{2}-i\delta_s)}.
\ee
\end{subequations}
Taking into account that the pump driving field is much more
intense than the signal field ($\vert \Omega_p \vert \gg \vert
\Omega_s \vert$), one obtains \cite{kim}  for the $\chi^{'}$
linear susceptibility and $\chi^{''}$ absorption rate \cite{houm}
\begin{subequations}
\label{susceptibility}
\bea
\label{susceptibility-a}
\chi^{'} &\approx& \varepsilon_{bac}+\frac{\Phi}{\hbar
\Xi} \, \gamma_{3} \left (\gamma_{3} \delta_s + \Omega_p \frac{\nu_2+\nu_3}{2} \right ) \nonumber \\
&+& \frac{\Phi}{\hbar \Xi} (\delta_p+\delta_s) \left (\nu_2\nu_3-\frac{\Omega^2_p}{4} +(\delta_s+\delta_p)\delta_s \right )
\eea
and
\be
\label{susceptibility-b}
\chi^{''} \approx \frac{\Phi}{\hbar \Xi} \gamma_{3} \left (\gamma_{3}\gamma_{2}+\frac{\Omega^2_p}{4} \right ) + \frac{\Phi}{\hbar \Xi}\left[\gamma_{2}(\delta_s+\delta_p)^2-\gamma_{3}\nu_2\nu_3\right],
\ee
\end{subequations}
where
$\Xi=[\gamma_{3}\gamma_{2}+\Omega^2_p/4-\nu_2\nu_3-(\delta_s+\delta_p)\delta_s]^2+[\gamma_{2}(\delta_s+\delta_p)+\gamma_{3}\delta_s+\Omega_p(\nu_2+\nu_3)/2]^2$,
$\Phi=\Gamma\vert \mu_{12}\vert
^2(\rho_{11}^{st}-\rho_{22}^{st})/\epsilon_0\Theta$,
$\varepsilon_{bac}$ is the background dielectric constant,
$\mu_{12}$ is the dipole moment of the transition between states
$\vert 1\rangle$ and $\vert 2\rangle$, $\epsilon_0$ is the vacuum
electric permittivity, and $\Theta$ is the volume of a single
quantum dot. $\Gamma$ is the optical confinement factor defined as
the fraction of the field intensity confined to the dots.
\cite{and} Here we assume that our structure is typical for
vertical-cavity quantum dot lasers, where light propagates
perpendicular to the active layer. Thus,
\[\Gamma\approx \zeta\frac{\int\limits_{dot} dz I(z)}{\int\limits_{structure} dz I(z)},\]
$I(z)$ being the field intensity as a function of the coordinate
along the direction of propagation and $\zeta$ is the ratio of
the dots area to the area of the structure in the direction
perpendicular to the signal-field propagation. Roughly, $\Gamma$
is proportional to the ratio of the total dot volume to the volume
of the structure. Here we assume that the dot remains on the lower
$\vert 1\rangle$ state, so that $\rho_{22}^{st}\ll\rho_{11}^{st}
\approx 1$. The slow-down factor, defined as the ratio of the
group velocities of light outside and inside the slowing medium,
is given in our case by
\begin{equation}
\Upsilon= n+\omega_s\frac{dn}{d\omega_s}, \label{slow}
\end{equation}
where $n=\sqrt{\chi^{'}+i\chi^{''}}$ is the complex refractive index.

\begin{figure}[t]
\begin{tabular}{c}
  \textbf{(a)} \\
 \includegraphics[scale=0.4]{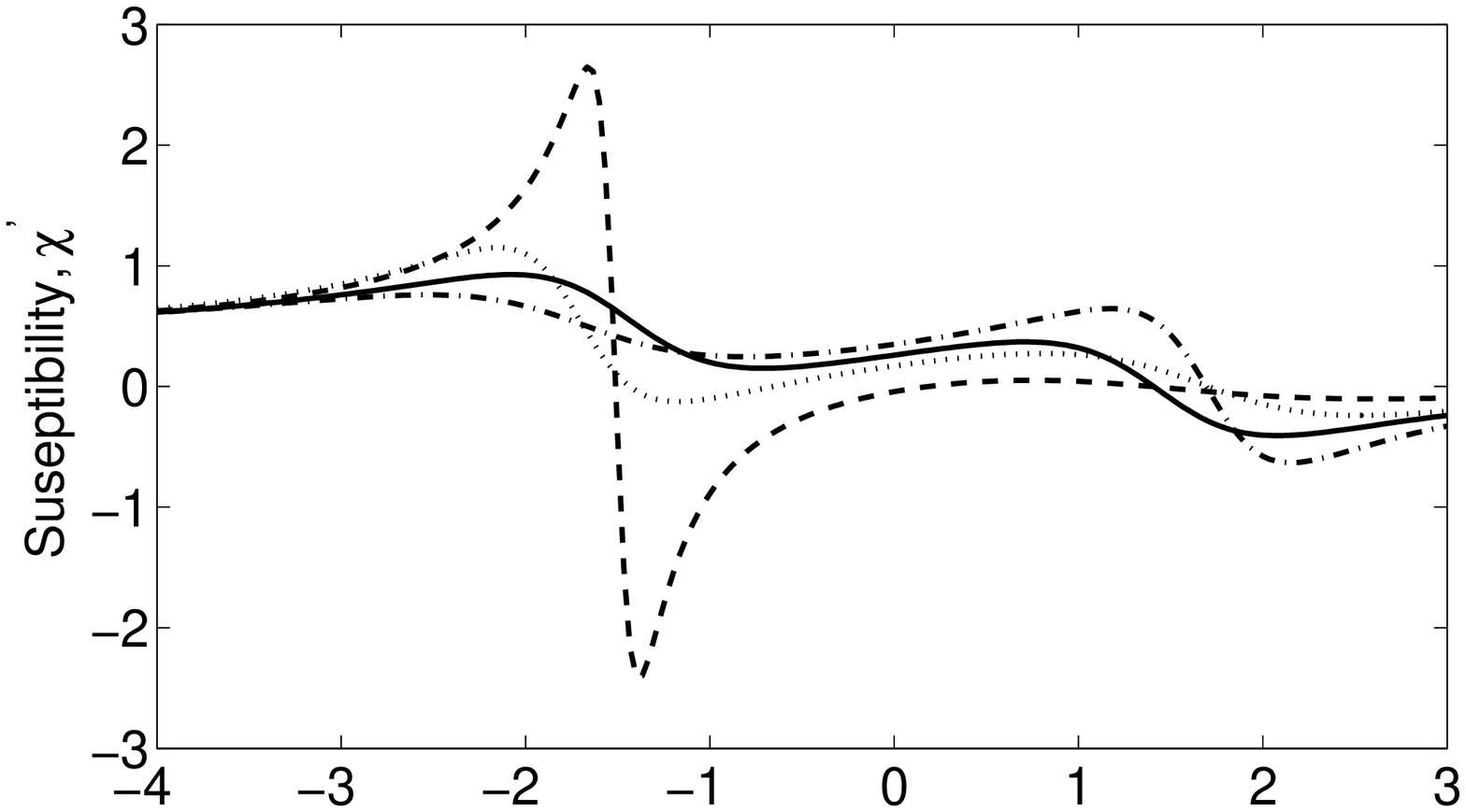}  \\
 \textbf{(b)} \\
  \includegraphics[scale=0.4]{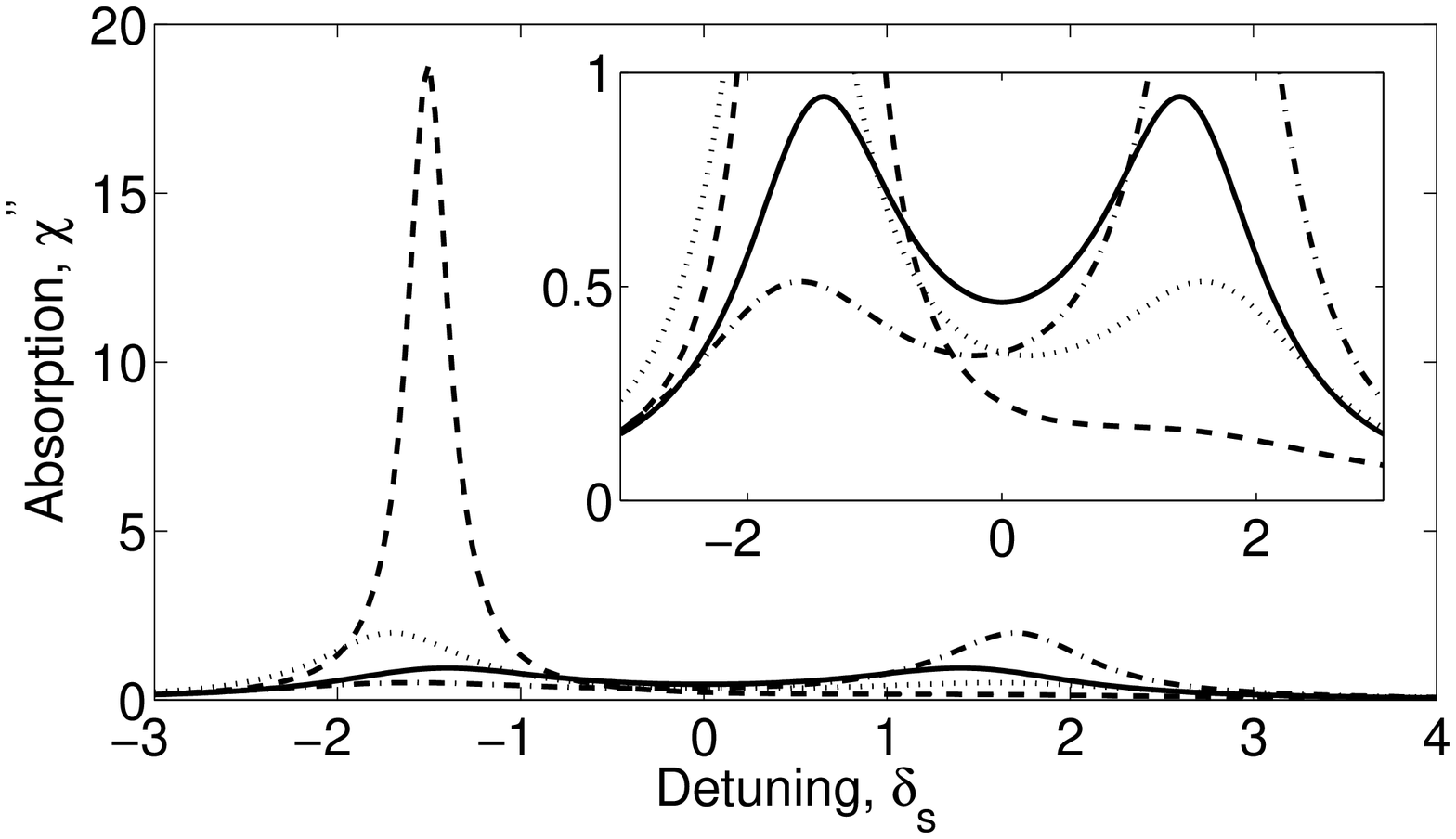} \\
\end{tabular}
\caption{(a) Normalized linear susceptibility and (b) absorption
profile in the weak non-Markovian case assuming $\nu_2\approx f_2 \, \Omega_p$ [$g_2 = g_3 = 0$ in Eq. (\ref{weak})]. Solid, dashed, dash-dotted and dotted  lines correspond to the $(f_2,f_3)=(0,0)$ Markovian case, $(f_2,f_3)=(-0.2,-0.4) \, \Omega_0$, $(f_2,f_3)=(-0.2,0.4) \, \Omega_0$, and $(f_2,f_3)=(0.2,-0.4) \, \Omega_0$, respectively. Other parameters are chosen as in Fig. \ref{fig3}. Inset in panel (b) shows the absorption profiles in more detail.} \label{fig5}
\end{figure}

\section{Non-Markovian effects}

Let us consider the asymptotic values of $\gamma_k$ and $\nu_k$ to establish the non-Markovian dynamics of the system. One then obtains
\bea
\label{g2ap}
\gamma_{2}&=&D^-_{22}(0)+\frac{s^2}{2}[D^-_{32}(0)-D^-_{22}(0)] - \frac{s^2}{4} \left [D^-_{32}(\Omega_R) \right. \nonumber \\
&-& D^-_{22}(\Omega_R) + \left. D^-_{32}(-\Omega_R)-D^-_{22}(-\Omega_R) \right ]
\eea
and
\bea
\label{g3ap}
\gamma_{3}&=&D^-_{33}(0)+\frac{s^2}{2}[D^-_{23}(0)-D^-_{33}(0)]- \frac{s^2}{4}\left [D^-_{23}(\Omega_R) \right. \nonumber \\
&-& \left. D^-_{33}(\Omega_R)+D^-_{23}(-\Omega_R)-D^-_{33}(-\Omega_R) \right ].
\eea
The $t \rightarrow \infty$ limit is equivalent to the Markovian approximation. We note that, due to the driving field, $\gamma_k$ and $\nu_k$ are dependent on the values of the Fourier-transforms [cf. Eq. \eqref{fourier cor }] of the reservoir correlation functions at $\delta=0,\, \pm \, \Omega_R$ (hence the name local Markovian approximation \cite{john}), whereas for the traditional Markovian approximation the rates are only dependent on the value of the Fourier-transforms at $\delta=0$. We assume weak non-Markovian effects, i.e., if $D_{kl}^-(\omega)$ is smooth in the vicinity of $\omega=0$ and varies only slightly on the scale defined by the modified Rabi frequency $\Omega_R$, then it may be represented as a polynomial up to second order in $\omega$. Therefore, Eq. (\ref{g2ap}) reduces to
\begin{eqnarray}
\label{g2ap reduced}
\gamma_{2} \approx \left. D^-_{22}(0)-\frac{\Omega_p^2}{4}\frac{d^2}{d\omega^2} \left [D^-_{32}(\omega)-D^-_{22}(\omega) \right ] \right \vert_{\omega\rightarrow 0}.
\end{eqnarray}
Similarly, one obtains for the $\gamma_3$ decay rate
\begin{eqnarray}
\label{g3ap reduced}
\gamma_{3} \approx \left. D^-_{33}(0)-\frac{\Omega_p^2}{4}\frac{d^2}{d\omega^2} \left [D^-_{23}(\omega)-D^-_{33}(\omega) \right ] \right \vert_{\omega\rightarrow 0}.
\end{eqnarray}
Therefore, the decay rates depend on the squared Rabi frequency associated with the driving field. Here we note that such a dependence leads to the damping of the driven Rabi oscillations. \cite{our prl} Similarly, for the coupling parameters from Eq. \eqref{fourier cor } one obtains
\bea
\label{nu2ap}
\nu_{2} &=& \frac{sc}{2} \left [D^-_{32}(0)-D^-_{22}(0) \right ]-\frac{s}{4}(1+c) \left [ D^-_{32}(\Omega_R) \right. \nonumber \\
&-& \left. D^-_{22}(\Omega_R) \right ] + \frac{s}{4}(1-c) \left [D^-_{32}(-\Omega_R) \right. \nonumber \\
&-& \left. D^-_{22}(-\Omega_R) \right ]
\eea
and
\bea
\label{nu3ap}
\nu_{3} &=& \frac{sc}{2}[D^-_{33}(0)-D^-_{23}(0)]+\frac{s}{4}(1-c) \left [D^-_{33}(\Omega_R) \right. \nonumber \\
&-& \left. D^-_{23}(\Omega_R) \right ] - \frac{s}{4}(1+c)\left [ D^-_{33}(-\Omega_R) \right. \nonumber \\
&-& \left. D^-_{32}(-\Omega_R) \right ].
\eea
As before, it is straightforward to show that Eqs. \eqref{nu2ap} and \eqref{nu3ap} lead to
\bea
\label{nu2ap reduced}
\nu_{2} &\approx& \left. \frac{\Omega_p}{2}\frac{d}{d\omega} \left [ D^-_{22}(\omega)-D^-_{32}(\omega) \right ] \right \vert_{\omega \rightarrow 0} \nonumber \\
&+& \left. \frac{\Omega_p\Delta_p}{2}\frac{d^2}{d\omega^2} \left [D^-_{22}(\omega)-D^-_{32}(\omega) \right ] \right \vert_{\omega \rightarrow 0}
\eea
and
\bea
\label{nu3ap reduced}
\nu_{3} &\approx& \left. \frac{\Omega_p}{2}\frac{d}{d\omega} \left [D^-_{33}(\omega)-D^-_{23}(\omega) \right ] \right \vert_{\omega\rightarrow
 0} \nonumber \\
&+& \left. \frac{\Omega_p\Delta_p}{2}\frac{d^2}{d\omega^2} \left [D^-_{33}(\omega)-D^-_{23}(\omega) \right ] \right \vert _{\omega\rightarrow 0}.
\eea
One may note that, for the calculation of the coherence coupling parameters, it is sufficient for the transforms of the $D_{kl}^-(\omega)$ bath correlation functions to be linearly dependent on the frequency to have the non-Markovianity affecting the susceptibility and absorption. As we shall see below, even for $\vert D_{kl}^-(\Omega_R)-D_{kl}^-(0) \vert \ll \vert \Omega_R \vert$, the influence of non-Markovianity on the susceptibility, absorption and the slow-down factor may be quite pronounced.

To demonstrate that it is quite common that non-Markovianity leads to significant coherence coupling parameters, let us consider a model of pure dephasing produced by the reservoir of acoustic phonons at low but finite temperature. Such a model has been extensively used for describing pure dephasing. \cite{forstner,rams,monni,ulrich,nazir,hkim,hug1,lodahl,hug2,calic,vuk} In the interaction picture with respect to the phonon reservoir, the reservoir operators are described by \cite{BP}
\be
\label{r boson}
R_k(t)=\sum\limits_{l}g_{kl}\left(b_le^{-iw_lt}+b_l^{\dagger}e^{iw_lt}\right),
\ee
where the $b_l$ and $b_l^{\dagger}$ are annihilation and creation operators of the phonon mode with frequency $w_l$ and the $g_{lk}$ are interaction constants (for simplicity, assumed as real).

As an example, let us take just one correlation function [i.e., $K_{22}(\tau,t)$] assuming there is no correlation between reservoirs. For the reservoir at temperature $T$,
\begin{widetext}
\begin{small}
\be
\label{k33}
K_{22}(\tau,t)=\sum\limits_{l} g_{2l}^2 \left [ \left [n_T(w_l)+1 \right ]e^{-iw_l(\tau-t)} + n_T(w_l) e^{iw_l(\tau-t)}\right ] = \int\limits_{0}^{\infty}dw J(w) \left [ \left [n_T(w)+1 \right ] e^{-iw_l(\tau-t)}+n_T(w)e^{iw_l(\tau-t)} \right ],
\ee
\end{small}
\end{widetext}
where the average number of thermal phonons in the mode is
\be
n_T(w)= \coth \left (\frac{\hbar w}{2k_BT} \right )
\ee
and the function $J(w)$ is the
density of states of the phonon reservoir. Let us consider a typical super-Ohmic density of states
\be
\label{density w3}
J(w)=\alpha w^3 \exp \left [ - \frac{w^2}{2w_c^2} \right ],
\ee
where $\alpha$ accounts for the interaction strength and $w_c$ is the cut-off frequency. Such a function describes a reservoir of acoustic phonons identified, for example, as a major source of pure dephasing in InGaAs/GaAs quantum dots. \cite{rams} We take $\alpha=0.4\pi^2$ ps$^{-2}$ (see, for example, Hughes \textit{et al.} \cite{hug2}) and assume a cut-off frequency $w_c=1$ meV. Fig. \ref{fig2} displays the dependence of the $\gamma_2$  driving-dependent decay rate [see Eq. (\ref{g2ap})] and $\nu_2$ coupling constant [cf. Eq. (\ref{nu2ap})] on the $\Omega_p$ Rabi frequency. We have scaled all the quantities with the  typical value \cite{kim} of the Rabi frequency of the driving field, $\Omega_0=6.6$ $\mu$eV. One may see that, for low temperatures (5 K-45 K was used for simulations in Fig. \ref{fig2}), the dependence of decoherence rates and coupling constants on the driving field is indeed quite pronounced and, quite definitely, may not be ignored.

\begin{figure}[t]
\begin{tabular}{c}
\includegraphics[scale=0.4]{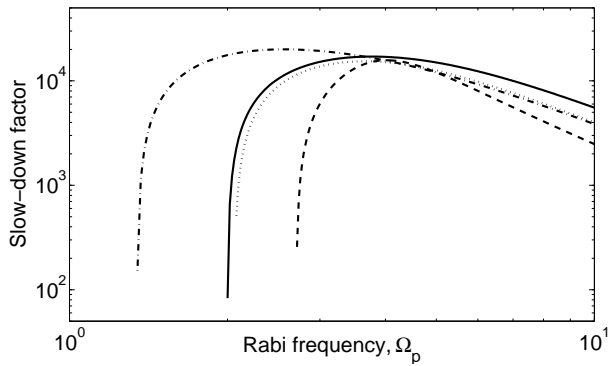}
\end{tabular}
\caption{Slow-down factor [cf. Eq.(\protect\ref{slow})] in the weak non-Markovian case assuming $\nu_2\approx f_2 \, \Omega_p$ [cf. Eq. (\ref{weak}) with $g_2 = g_3 = 0$] and $\delta_p = \delta_s=0$. Solid, dashed, dash-dotted and dotted lines correspond to the $(f_2,f_3)=(0,0)$ Markovian case, $(f_2,f_3)=(-0.2,-0.4) \, \Omega_0$, $(f_2,f_3)=(-0.2,0.4) \, \Omega_0$, and $(f_2,f_3)=(0.2,-0.4) \, \Omega_0$, respectively. Other parameters are chosen as in Fig. \ref{fig3}.} \label{fig6}
\end{figure}

\begin{figure}[t]
\includegraphics[scale=0.4]{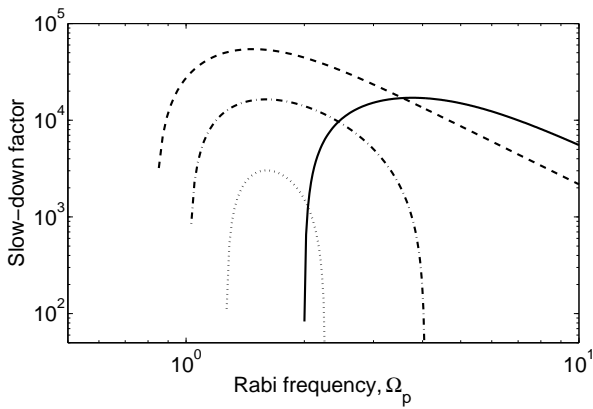}
\caption{Slow-down factor in the weak non-Markovian case. Here [cf. Eq. \eqref{weak}] $f_3=2 \, f_2$, $g_3=2 \, g_2$, $\delta_p=\delta_s=0$. Solid, dashed, dash-dotted and dotted lines correspond to the $(f_2,g_2) = (0,0)$ Markovian case, $(f_2,g_2)  =(0.2,0.0001) \, \Omega_0$, $(f_2,g_2)=(0.2,0.1) \, \Omega_0$, and $(f_2,g_2) = (0.2,0.15) \, \Omega_0$, respectively. Other parameters are chosen as in Fig. \ref{fig3}.} \label{fig7}
\end{figure}

\section{Discussion and Examples}

Here we wish to investigate the influence of non-Markovian effects on the susceptibility and absorption of a three-level quantum dot.  To this end, we compare the Markovian and non-Markovian regimes and analyze the influence of non-Markovian effects and correlated dephasing reservoirs.

In the Markovian limit we choose the values of parameters used by Kim \textit{et al.} \cite{kim} in the case of a cylindrical strained GaAs-InGaAs-InAs quantum-dot system, i.e.,  $\varepsilon_{bac}=13$, $\vert \mu_{21}\vert /e=2.1$ nm, $\Gamma=0.006$, $V=8.9064 \times  10^{2}$ nm$^3$, and the wavelength of the signal field is 1.36 $\mu$m, which is much shorter than the wavelength of the driving field (12.8 $\mu$m). Also, it was assumed that the dephasing rate for level 3 was two times larger than for level 2. In the results shown below we also make this assumption for both the dephasing rates and coupling parameters $\gamma_3=2\gamma_2$ and $\vert \nu_3 \vert=2 \vert \nu_2 \vert$. Moreover, here we assume that decay rates and coupling constants depend on the Rabi frequency associated with the driving field [cf. Eqs. \eqref{g2ap reduced}-\eqref{g3ap reduced} and \eqref{nu2ap reduced}-\eqref{nu3ap reduced}] as
\bea
\label{weak}
\gamma_k \approx \gamma_k^{(0)} + \frac{g_k}{2} \Omega_p^2, \quad \nu_k \approx
f_k \Omega_p + g_k \Delta_p \Omega_p,
\eea
for $k = 2, \,3$. As mentioned, $D_{kk}^-$ has been represented as a polynomial up to second order in $\omega$, i.e., $D_{kk}^- (\omega) \approx \gamma^{(0)}_k + 2 f_k \, \omega + g_k \, \omega^2$ for $k = 2, \,3$. In addition, $D_{32}^- (\omega) = D_{23}^- (\omega) = 0$. It should be noted here that $\gamma_k^{(0)}$ rates might also include a contribution from Markovian reservoirs describing energy loss and other sources of Markovian dephasing. Let us first consider the case of a negligible change of the decay rate with the Rabi frequency, i.e., $\gamma_k^{(0)} \gg \vert \frac{g_k}{2} \Omega_p^2 \vert$ for $k = 2, \,3$. Weak non-Markovian effects ($\vert f_k \Omega_p\vert \ll \gamma_k^{(0)}$, $k = 2, \,3$) on the susceptibility, absorption and slow-down factor, are then illustrated in Figs. \ref{fig3} and \ref{fig4}. Fig. \ref{fig3}(a) depicts the susceptibility for the Markovian resonant ($f_2 = 0$, $\Delta_p=0$) and two non-Markovian cases ($f_2 = 0.1 \, \Omega_0$, $f_2 = 0.2 \, \Omega_0$, $\Delta_p=0$) as a function of the signal-field $\delta_s$ detuning. One notices that the slope of the non-Markovian susceptibility curve is indeed rather pronouncedly enhanced. In addition, as indicated by results from the absorption profiles in Fig. \ref{fig3}(b), non-Markovian effects shift and narrow the transmission bandwidths (the latter is natural to expect; see, for example, Tucker \textit{et al.} \cite{buffers}). Moreover, the increase in the slow-down factor may be quite large (cf.  Fig. \ref{fig4}) whereas the transmission bandwidth is not significantly reduced. Non-Markovian effects are phase-sensitive and may be either harmful or advantageous for slowing down light. Simultaneous changes of sign of coupling constants $\nu_2$ and $\nu_3$ shift the peaks of both the susceptibility and absorption to the region of negative $\delta_s$ detunings (see Figs. \ref{fig5} and \ref{fig6}). However, the change is not completely symmetrical and gain in the slow-down factor is smaller than in the case of $\nu_2$ and $\nu_3$ both positive. It is interesting to point out that taking $\nu_2$ and $\nu_3$ of opposite signs, one may simultaneously obtain both the enhancement of the slow-down factor and widening of transmission bandwidth  [see inset in Fig. \ref{fig5}(b) and Fig. \ref{fig6}].

Up to now we have considered decay rates independent of the Rabi frequency $\Omega_p$. As we have shown before, such dependencies may be considered through Eq. \eqref{weak} by taken $g_k \neq 0$ ($k=2, \,3$). In this case, the non-Markovian nature of the EIT process may destroy the efficiency in obtaining slow light. To see this, we turn to Fig. \ref{fig7} where the behavior of the slow-down factor is plotted as a function of the Rabi frequency associated with the pump field. If decay rates depend on the Rabi frequency $\Omega_p$, one may obtain EIT systems with worse performance than in the Markovian case. As such dependence becomes more remarkable, i.e, as $g_2$ and $g_3$ are increased [cf. Eq. \eqref{weak}],  a shifting and narrowing of the transmission window is more noticeable, and a complete loss of the non-Markovian slow-down effect may be obtained as the Rabi frequency $\Omega_p$ is increased.

We note that correlations between reservoirs may affect the influence of memory effects on the slow-down factor. In a quantum dot system, the interaction between the dot electrons and its surrounding medium, i.e., impurities, phonons, etc., may lead to significant correlations between reservoirs. Moreover, by construction, operators $R_k(t)$ ($k=2, \, 3$) contain variables corresponding to the same reservoir acting on the lower level of the dot. To see how the correlation between reservoirs is inevitably arising in conventional models of dephasing, let us consider an example of the electron-phonon interaction constants [cf. Eq. \eqref{r boson}] for the present three-level dot. They may be represented in the following general
form \cite{hug1}
\be
\label{intc}
g_{kl}=C\sqrt{\hbar \vert \bvec{L} \vert}\int d^3 \bvec{r} \left [d_k \vert \phi_k(\bvec{r})\vert ^2-d_1 \vert \phi_1(\bvec{r}) \vert^2 \right ]e^{i\bvec{L}\cdot \bvec{r}},
\ee
where $\bvec{L}$ is the wave-vector of the $l$-th phonon mode, $d_k$ and $\phi_k(\bvec{r})$ are the deformation potential and the electron wave-function of the $k$-th state of the dot, respectively, and $C$ is a constant. Since electron wave-functions are localized in the vicinity of the dot, and taking into account the small energies of acoustic phonons (in the present case it would be much less than 1 meV), the densities of states of the phononic reservoirs corresponding to states $\vert 2\rangle$ and $\vert 3\rangle$ should be similar. The difference in the integrand of Eq. \eqref{intc} involves essentially the values of the $d_k$ deformation potential.  If the deformation potentials are close in values for different excited states, the model defined in Eq. \eqref{r boson} predicts obliteration of the non-Markovian effect by correlated dephasing. Notice that, for an adequate treatment of correlations between reservoirs, one needs to appropriately account for phase differences between constants $g_{kl}$ to develop a realistic microscopic dephasing model.

\section{Conclusions}

In conclusion, we have shown that it is necessary to take into account non-Markovian effects when considering slowing down light in EIT schemes based on quantum dots. Non-Markovian behavior is typical for such systems, especially in low temperature conditions. We have demonstrated that, for decay rates independent of the Rabi frequency associated with the pump field, even relatively weak non-Markovian effects may lead to significant enhancement of the slow-down factor together with the simultaneous broadening of the transmission window. However, if decay rates are considered to be dependent of the Rabi frequency $\Omega_p$, a shifting and narrowing of the transmission window may be obtained. Furthermore, non-Markovian effects may lead to significant driving-induced dephasing which may inhibit the slowing-down effect. Moreover, it is suggested that the presence of correlation between reservoirs may remove the harmful effects produced by the non-Markovian nature of EIT. Finally, considering the importance of investigations to produce efficient slow light propagation in solid-state systems, we do hope this work would stimulate future experimental and theoretical studies on this subject.

\acknowledgments

This work was partially supported by the National Academy of
Sciences of Belarus through the program ``Convergence", and by
FAPESP grant 2014/21188-0 (D. M.). The authors would also like to
thank the Scientific Colombian Agencies Estrategia de
Sostenibilidad (2015-2016) and CODI - University of Antioquia, and
Brazilian Agencies CNPq, FAPESP (Proc. 2012/51691-0) and
FAEPEX-UNICAMP for partial financial support.

\end{document}